\documentclass[preprint]{elsarticle}

\usepackage[utf8]{inputenc}  
\usepackage[T1]{fontenc}     
\usepackage{lmodern}         
\usepackage{microtype}       
\usepackage[scaled]{helvet} 
\usepackage[english]{babel}  
\usepackage{graphicx}        
\usepackage{hyperref}
\hypersetup{bookmarks=false}
\usepackage{fullpage}
\usepackage{comment}         
\biboptions{sort&compress}   

\journal{arXiv}

\usepackage{amsmath,amsfonts,amssymb}

\usepackage{subfigure}

\usepackage{xcolor}
\usepackage{todonotes}
\usepackage{multicol}

\usepackage{caption}

\graphicspath{{fig/}}

\begin{document}

\begin{frontmatter}


\title{New GO-based Measures and Their Statistical Significance in Multiple Network Alignment}

\author{Reza Mousapour\fnref{label1}%
}
\author{Kimia Yazdani\fnref{label2}%
}
\author{Wayne B. Hayes\fnref{label2}%
}
\fntext[label1]{Department of Computer Engineering, Sharif University of Technology, Tehran, Iran}
\fntext[label2]{Department of Computer Science, University of California, Irvine, CA 92697-3435, USA.}

\begin{abstract}
 Protein-protein interaction (PPI) networks provide valuable insights into the function of biological systems, and aligning multiple PPI networks can reveal important functional relationships between different species. However, assessing the quality of multiple network alignments is a challenging problem. In this paper, we propose two new measures, the Squared GO Score (SGS) and the Exposed G Score, to evaluate the quality of multiple network alignments while using functional information from Gene Ontology (GO) terms. We also introduce a $p$-value measure, the Statistical Exposed G Score, to compute the exact significance of a multiple network alignment based on its revealed GO terms. We also show that our measures are highly correlated with the recovered Ortholog count, providing further evidence for their effectiveness. Our work contributes to the development of more reliable and accurate measures for evaluating multiple network alignments and has potential applications in predicting gene function and identifying evolutionary relationships between different species using multiple network alignment.

\vspace{1\baselineskip}


\end{abstract}

\end{frontmatter}


\section{Introduction} 
Protein-Protein Interaction (PPI) networks have become a popular tool for investigating biological systems, as they show the interactions between proteins and the functions they perform. PPI networks can be used to predict gene ontology (GO) terms, protein function, and identify potential drug targets. 

However, these networks are often noisy and incomplete, which can limit their utility \cite{predictpro}. Network alignment methods have the potential to overcome these limitations. The alignment process seeks to identify conserved regions across networks, and to maximize topological similiarity between the networks. However, evaluating the quality of multiple network alignments remains a challenge.

Several measures have been proposed for evaluating the quality of multiple network alignment. One commonly used measure is the Maximal Common Subgraph (MCS), which quantifies the size of the largest common subgraph between the aligned networks \cite{Bunke98}. Another measure is the Normalized Compression Distance (NCD), which evaluates the similarity between the compressed representations of the aligned networks \cite{cilibrasi2005clustering}. Both these measures are topology-based and they provide information only about the structural similarity between the aligned networks, without considering the functional aspects of the networks. Although it is widely {\it assumed} that common topology relates closely to common function, this has yet to be observed or objectively measured. As a result, in biological network alignment, it is crucial to consider the functional information encoded in GO terms, which offers important information about the biological roles and relationships of the aligned proteins \cite{GO}.

The use of GO terms has been proposed as a way to assess the quality of pairwise and multiple network alignment methods beyond topology-based measures. One such method is the Functional Overlap (FO) score, which computes the ratio of shared GO terms between aligned nodes in the multiple network alignment \cite{Gligorijevic2016}. FO only uses the ratio of shared GO terms and do not take into account the quality of the alignment for each individual GO term. Measures like FO assume equal importance for different GO terms, whereas some of GO terms can provide more useful information, for instance due to their lower frequency. Therefore, it is important to look at each GO term in its own context and evaluate them separately. 

Our work aims to address the mentioned challenges by proposing measures to compute the exact $p$-value of a multiple network alignment with respect to each GO term separately, and then combining their results to produce a single holistic $p$-value across all GO terms. To our knowledge, this is the first work that evaluates the statistical significance of multiple network alignments by calculating a $p$-value for each alignment. Additionally, we introduced two function-based quality measures for assessing the quality of multiple network alignment based on a single GO term. Our measures provide a comprehensive framework for evaluating the quality of multiple network alignments, which can help researchers to identify the most reliably aligned regions and improve the accuracy of their predictions and facilitate the identification of new functional relationships between proteins.

Moreover, by determining the statistical significance of a multiple network alignment, we can assess the reliability of the alignment and determine the likelihood of obtaining the same alignment by chance. The significance of obtaining an alignment or a better one, can be used as a rigorous indicator of quality. A key difference of this measure is that it can be used to compare two multiple network alignments that have been created between different set of species. By providing measures to evaluate the quality and statistical significance of these alignments, researchers can make more informed decisions and drive new discoveries in the field. Although the measures presented in this study were applied to biological PPI networks, they can be extended to any type of multiple network alignments that are annotated with an ontology, including social networks and other complex systems.


\section{Materials and Methods} 
\subsection{Definitions}
We consider multiple network alignment in a 1-to-1 sense, where each node is allowed to be in only one cluster, but every node must be in a cluster. For $k$ undirected graphs $G_i, i=1,\ldots,k$, let $G_i$ have $n_i$ nodes and $\lambda_i$ nodes annotated with GO term $g$. The binomial coefficient $c(n,x)$ and the number of permutations $p(n,x)$ are used, where $n$ is the total number of objects and $x$ is the number of selected objects.
Without loss of generality, we assume two conventions for sorting the networks: For \textbf{node-sorted} components, where the networks are sorted by node size, we assume: $n_1 \geq n_2 \geq ... \geq n_k$
For the \textbf{lambda-sorted} components, where the networks are sorted by the density of $g$ we assume: $\lambda_1 \geq  \lambda_2 \geq, ... \geq \lambda_k$.
We also employ a {\it Shadow Network} \cite{sana}, which schematically depicts the state of a multiple network alignment. Each node in the shadow network represents a cluster of aligned nodes, and edges between shadow notes carry an integer weight representing the number of aligned edges between two clusters. The $k$ networks that are being aligned are considered peg-networks and the shadow network is considered a hole-network that can be larger than all of the networks. Although the equations allow for extra shadow nodes, for practical cases, the maximum size of the hole-network considered for quality measures is assumed to be no more than the size of the largest peg-network. We note that the existence of shadow nodes in the hole-network, which do not correspond to any node in a peg-network, should not affect the quality measures. 

Overall, our proposed conventions and measures aim to provide a consistent and comparable framework for evaluating multiple network alignment using a single GO annotation term $g$.

\subsection{Equations}
\paragraph{Preliminaries:} We represent a cluster of mutually aligned nodes as a {\it tower} of pegs. Recall that, for now, we are working with a single GO term, called $g$. Thus, each protein in the cluster (tower) is either annotated by $g$, or not. Given tower $T$, let $a_T$ be the number of proteins in $T$ that are annotated by $g$. We propose two measures for evaluating a multiple network alignment: SGS, and EG.

Note that these measures are never used to {\em guide} the alignment, but are only used after-the-fact to evaluate an alignment generated without their use. The general principle is that an alignment that that “concentrates” more GO terms into fewer towers is better than spreading GO terms around equally between towers. This concentration can be measured in two ways: we can either reward towers that have more GO terms than average (the SGS score below), or we can reward the {\em entire alignment} if all the GO terms are concentrated into fewer towers (the “exposed GO” score below). Thus, better alignments have a larger SGS score, but a {\em smaller} EG value. EG is used in the denominator in the fnial equation of EG score, so the EG score is also higher if we have a better alignment.
\subsubsection{Squared G Score (SGS)}
The proposed SGS score is a measure for evaluating the quality of multiple network alignments. The components are assumed to be lambda-sorted for this part. This measure  computes the sum of the squared number of $g$-annotated nodes in each tower. The higher the SGS score, the better the alignment is. We define the SGS score for tower $T$ as $|a_T|^2$, the square of the number of annotated proteins. Then, the global score of an alignment $F$ is

\begin{equation}
    SG(F)=\sum_T |a_T|^2. 
    \label{eq:1}
\end{equation}

\noindent to normalize the SGS score between 0 and 1, we divide it by its maximum possible value, which is computed as follows:

\begin{equation}
    Max(SG) = k^2\lambda_k + \Sigma_{i=1}^{k-1}i^2(\lambda_i - \lambda_{i+1})
    \label{eq:1}
\end{equation}

\begin{equation}
    SGS = \frac{SG}{Max(SG)}
    \label{eq:2}
\end{equation}

\noindent where $k$ is the number of networks in the alignment.

\subsubsection{Exposed G Score}
Exposed $G$ is another measure of the quality of a multiple network alignment. Here, we want to reward alignments where all the GO terms are concentrated into a smaller number of towers. We imagine that GO annotations in a tower are aligned vertically, and the uppermost GO term in the tower is “expose”, while the ones “below” it are not. Thus, a better alignment is one that has a smaller number of exposed GO terms. A lower exposed $G$ score indicates that more $g$-annotated nodes are aligned together and fewer nodes are "exposed," resulting in a better alignment. For this score too, the components are lambda-sorted. 

\begin{equation}
    \text{Exposed g = Number of the towers with at least one $g$ annotated node}
    \label{eq:2}
\end{equation}

The calculation of exposed $G$ involves sorting the components by $\lambda$ and counting the number of towers with at least one $g$-annotated node. The minimum value of exposed $G$ is $\lambda_1$ and the maximum value is $\Sigma_{i=1}^{k}\lambda_i$, where $i$ is the index of the network and $k$ is the total number of networks. The exposed $G$ score is then calculated as the minimum value of exposed $G$ divided by the actual exposed $G$ value.

\begin{equation}
    \text{Min(Exposed g)= } \lambda_1
    \label{eq:2}
\end{equation}

\begin{equation}
    \text{Exposed g Score =}\frac{\text{Min(Exposed g)}}{\text{Exposed g}}
    \label{eq:2}
\end{equation}

To incorporate exposed $G$ into the final alignment score, it is placed in the denominator. This transforms it from a cost measure into a score measure between 0 and 1. The simplicity and effectiveness of exposed $G$ make it a valuable addition to the set of measures used to evaluate multiple network alignments.

\subsubsection{Statistical Exposed G, the $p$-value of a multiple network alignment}

Scores like SGS and EG are heuristic; they are not rigorous in the sense that we have no idea how “good” a particular score is compared to a random alignment. Here, we wish to transform the Exposed G score into a rigorous $p$-values. In order to evaluate the statistical significance of a multiple network alignment, we utilize exposed $g$ as a measure to calculate the $p$-value of a specific alignment. The denominator of the $p$-value calculation represents the total number of possible ways a multiple network alignment can be created, while the numerator counts the number of multiple network alignments with a particular exposed $g$ value.

By determining the exposed $g$ value of a given multiple network alignment and utilizing the equations for the numerator and denominator, we can calculate the statistical significance of obtaining that specific exposed $g$ value. This approach allows for a more comprehensive evaluation of the quality of a multiple network alignment, as the statistical significance of obtaining a specific exposed $g$ value or a better (lower) one can serve as a reflection of its quality. By calculating the $p$-value of getting a certain exposed $G$ value or lower, we can assess how likely it is to observe such an alignment by chance, which can reflect the quality of the alignment in a more rigorous and objective manner.

It should be noted that in this context, exposed $g$ refers to the actual number of exposed GO terms, rather than the normalized exposed $g$ score. Furthermore, the calculation of the numerator and denominator is described in two separate sections to provide a clear understanding of the statistical significance calculation for a single GO term.

\renewcommand{\paragraph}[1]{\smallskip\noindent\textbf{#1}}
\paragraph{2.2.3.a The denominator}

The denominator in this context refers to the combinatorial number of ways that a multiple network alignment can be created between a set of networks, taking into consideration the existence of shadow nodes with no biological meaning. The ordering of these shadow nodes does not matter and their number is typically set to zero, but the general case is considered where the number of extra shadow nodes is bound to be between $n_1$ to the sum of network sizes. It is important to note that this is the whole possible range, but usually the user defines an upper bound that is much closer to $n_1$. We refer to this defines upper bound as $n_0$. To calculate the denominator, we assume that the networks are node-sorted and each network is independently matched with the shadow network. 

\begin{equation}
    LE(n) = \Pi_{i=1}^{k}p(n, n_i)
    \label{eq:2}
\end{equation}

\noindent where $n=n_0$, and $n_0$ is the number of shadow nodes allowed for the network. This equation is also the final equation that can be used when we set the extra number of shadow nodes to zero.

However, given that number of shadow nodes is defined as $n_0>n_1$, there are a number of empty shadow nodes in each alignment, and different orderings of filled shadow nodes are incorrectly accounted for. Therefore, $LE(n)$ counts the number of ways that $n$ or fewer shadow holes are filled, allowing for repetitive counts for each case. The exact number of ways that exactly $n$ shadow holes are filled is represented by $E(n)$, which is defined as

\begin{equation}
    LE(n) = \Sigma_{i=n_1}^{n}a_i E(n_i)
    \label{eq:2}
\end{equation}

\noindent where $a_i$ represents the factor of $E(n_i)$ being counted in $LE(n)$, where generally, $a_i=p(n,i)$ since for the shadow network, $LE$ chooses $i$ shadow holes and permutes them. 

\begin{equation}
    LE(n) = \Sigma_{i=n_1}^{n}\frac{n! E(n)}{(n - i)!}
    \implies E(n) = \Sigma_{i=n_1}^{n}\frac{LE(i)\times (-1)^{n-i}}{(n-i)!i!}
    \label{eq:2}
\end{equation}

Finally, the count of different multiple network alignments is represented by

\begin{align}
    \text{Denominator} &= \Sigma_{n=n_1}^{n_0}E(n) = \Sigma_{n=n_1}^{n_0}(\Sigma_{i=0}^{n_0 - n}\frac{(-1)^i}{i!})LE(i) \\
    &= \Sigma_{n=n_1}^{n_0}(\Sigma_{i=0}^{n_0 - n}\frac{(-1)^i}{i!(n-i)!})\Pi_{i=1}^{k}p(i,n_i)
\end{align}

This equation takes into consideration the number of shadow holes filled and the factorials of each combination. The denominator essentially represents the total number of ways that multiple network alignments can be formed between the given set of networks.

\renewcommand{\paragraph}[1]{\smallskip\noindent\textbf{#1}}
\paragraph{2.2.3.b The numerator}

Here we want to calculate the exact number of alignments that have $x$ exposed $g$s. Again, we define $LEG(x)$ to be the number of ways that we have $x$ or less exposed $g$s. There are some duplicates in $LEG(x)$ that we should eliminate by defining an exact version. 
To evaluate the quality of a multiple network alignment, we want to calculate the exact number of alignments that have a given number of exposed $g$ values. We define $LEG(x)$ to be the number of ways that we have $x$ or fewer exposed $g$s. However, there are duplicates in $LEG(x)$ that we must eliminate. Therefore, we define $EG(x)$ to be the exact number of alignments that have exactly $x$ exposed $g$s. Since we know that in the $\lambda$-sorted networks, the lower bound for exposed $g$ is $\lambda_1$, we can rewrite $LEG(x)$ as:

\begin{equation}
    LEG(n, x) = c(n, x)\Pi_{i=1}^{k}p(x, \lambda_i)p(n-\lambda_i, n_i - \lambda_i)
    \label{eq:2}
\end{equation}

In the equation, we first choose $x$ nodes from the shadow nodes and then for each of the networks, we choose all of the annotated nodes among these $x$ nodes. Then, we align the rest of the network nodes with the remaining shadow nodes.  In cases where all $x$ chosen shadow nodes are aligned with at least one annotated node, there will be no duplicate counting. However, in other cases, there are duplicates because the shadow nodes that are not aligned with any annotated nodes are the same, whether they are chosen from $x$ or not. The annotated nodes in $x$ are not counted duplicated because they can only be in $x$ and it matters to which shadow nodes they are aligned. But the unannotated ones could as well be left out of $x$. As a result, we define $EG(x)$ to be the exact number of alignments that have $x$ exposed $g$s. Since we know that in the $\lambda$-sorted networks, the lower bound for exposed $g$ is  $\lambda_1$, we rewrite $LEG(x)$ as

\begin{equation}
    LEG(x) = \Sigma_{i=0}^{x - \lambda_1}a_i EG(i)
    \label{eq:2}
\end{equation}

Here, $a_i$ is the factor being counted in $EG(i)$. For $i$ exposed $g$s, $a_i=c(n-(x-i),i)$, where $c(n,k)$ is the number of ways to choose $k$ items from a set of $n$ items. The equation for $LEG(x)$ involves choosing $x$ nodes from the shadow nodes and, for each network, choosing all of the annotated nodes among these $x$ nodes. Then, we align the rest of the network nodes with the remaining shadow nodes. Since

\begin{equation}
    LEG(n,x)= \Sigma_{i=0}^{x-\lambda_1}c(n-(x-i),i)EG(x-i) 
    \label{eq:2}
\end{equation}

We can calculate $EG(n,x)$ as

\begin{equation}
    EG(n,x)=\Sigma_{i=0}^{x-\lambda_1}LEG(n,x-i)c(n-(x-i),i)(-1)^i
    \label{eq:2}
\end{equation}

Here, we also need to solve the shadow hole discrepancy. For this part, we can use the denominator equations since the equations have the same number of undercounting for different cases.

\begin{equation}
    E(x)=\Sigma_{n=n1}^{n0}\Sigma_{i=max(n1,x)}^{n0-n}\frac{EG(n,i)\times(-1)^{n-i}}{(n-i)!i!}
    \label{eq:2}
\end{equation}

In order to calculate the statistical significance of this, we should calculate $E(x)$ for Exposed $g$ s equal or better than the counted Exposed $g$. So, the statistical significance of obtaining an exposed $G$, can be calculated by $E(x)$, but in order to calculate the statistical significance of obtaining an exposed $G$ or a lower one, we should use this equation:

\begin{equation}
    Numerator=\Sigma_{i=\lambda_1}^{min(sum(lambdas),n_0)}E(x)
    \label{eq:2}
\end{equation}

This approach allows for a more comprehensive evaluation of the quality of a multiple network alignment, as the $p$-value of obtaining a specific exposed $g$ value or a better (lower) one can serve as a reflection of its quality.

\subsection{Validation and Logarithmic Implementation of Statistical Exposed G}

The statistical exposed G measure has been proposed as a means to evaluate the degree of conservation of a set of genes across multiple organisms. Although the equations for the statistical exposed G measure can be difficult to follow, we have validated the measure through numerical and empirical methods.

To validate the measure numerically, we generated random network states and compared the sum of the numerator for all possible values of exposed Gs with the denominator. We then implemented the functions by logarithmic calculations to allow for the calculation of results for larger values. We tested the output for different network states, including real-sized and smaller networks, and observed that the result was equal for the numerator and the denominator. We only needed to produce different lambda values lower than the network sizes, an arbitrary allowed shadow network size in the possible range. For the logarithmic calculations, we calculated $log(a+b)$ having $log(a)$ and $log(b)$ as in the equation

\begin{equation}
    log(a+b)=log(1+(exp(log(a)-log(b))) + log(b)
    \label{eq:2}
\end{equation}

\noindent assuming $b$ is larger than a. For a value of $exp(log(a)-log(b))$ close to zero, we used the Taylor series. Additionally, we had two sets of series that were multiplied by different powers of -1. Consequently, we also needed to calculate $log(a-b)$ which was done by factoring $b$ as demonstrated in the equation

\begin{equation}
    log(a-b)=log(1-exp(log(b)-log(a)))+log(a)
    \label{eq:2}
\end{equation}

\noindent again Taylor series was used for small values of $exp(log(b)-log(a))$.

To further validate the statistical exposed G measure, we used empirical tests. We generated numerous random alignments, calculated the exposed G, and tracked the number of times each exposed G was observed for a network state (which includes $n_1$ to $n_k$ and $\lambda_1$ to $\lambda_k$). Dividing this number by the number of times that this network state was produced gave us the “empirical $p$-value” of the specific exposed G for the specific network state. We then used our equation to calculate the “theoretical $p$-value” for each of these values. We used IID mammalian networks for this purpose, with Rat, Cow, and Dog networks used for the case with three networks and 100 million random alignments generated. For the case with 5 networks, 9 billion random alignments were generated.

It is important to note that if the theoretical $p$-values are much lower than $10^{-9}$, the comparisons would be invalid since the sample size for the empirical results wouldn’t be enough to represent them. In multiple network alignment, due to the numerous alignment states, the $p$-values generally decrease very fast. Therefore, we chose a subset of 100 nodes for each of these networks to get theoretical results within this range. Our method for choosing a hundred nodes included going through the Orthologs between the species and choosing the 100 sets of orthologs that had the highest minimum number of GO terms, selecting nodes with the highest shared GO terms.

\begin{figure}[htbp]
  \centering
  \includegraphics[width=1\textwidth]{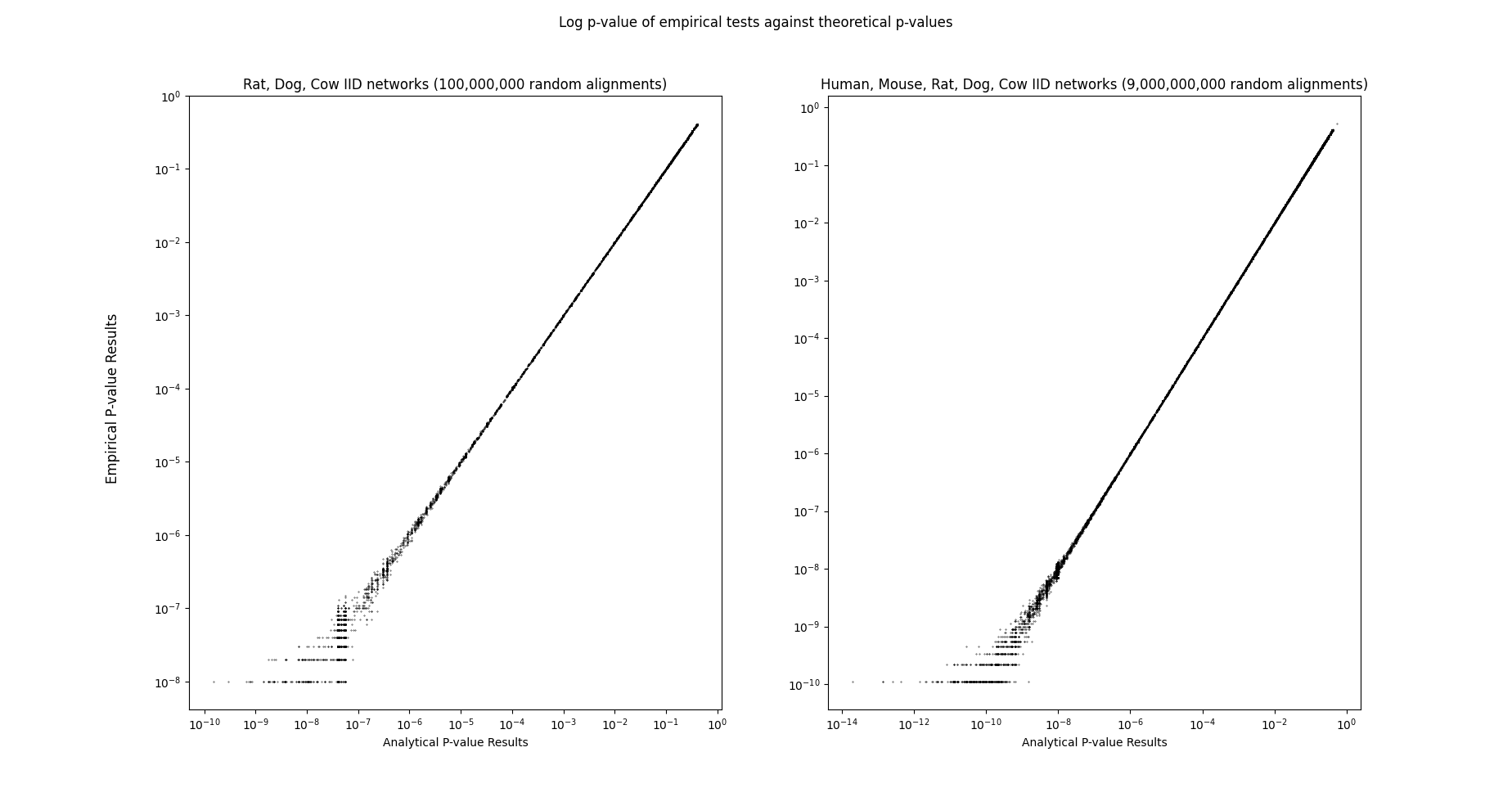}
  \caption{The scatterplot of empirical $p$-values for statistical exposed G against the theoretical values. The left plot shows the results for three networks and 100 million randomly generated alignments. The right plot demonstrates the validation results for five IID mammalian networks and 9 billion randomly generated alignments.}
  \label{fig:empiricaltest}
\end{figure}

The empirical tests show excellent agreement with the theoretical results as shown in figure \ref{fig:empiricaltest}. The divergence in the lower $p$-values is due to the relatively small sample size that is not representative enough for these groups.

\subsection{Experiments}
Two main sets of experiments were used to assess the measures. These experiments compare the measures against an objective evidence to test their quality.   
\subsubsection{Perfect self-alignment with controlled error rate}
We used BioGrid networks for the first experiment \cite{chatr2017biogrid}. Our goal was to produce controlled noise using self-alignment. To achieve this, we selected human, mouse, rat, and fruit fly networks. In each set, we began with a perfect self-alignment of $k$ numbers in a single-species network. Next, we permuted a specific fraction of nodes in the networks to introduce a predetermined error rate. We generated 250 alignments for each combination of k, error rate, and species, resulting in a total of 30,000 self-alignments. A summary of the utilized BioGrid networks are demonstrated in the table \ref{tab:biogrid}.
\begin{table}[ht]
\centering
\caption{A summary of experimented BioGrid networks}
\label{tab:biogrid}
\begin{tabular}{ccccc}
\hline
Nodes & Common Name & Official Name & Abbr & TaxID\\
\hline
13276 & Human & H. Sapiens & HS & 9606 \\
7937 & Fruit fly & D. Melanogaster & DM & 7227\\
4370 & Mouse & M. Musculus & MM & 10090\\
1657 & Rat & R. Norvegicus & RN & 10116\\
\hline
\end{tabular}
\end{table}

\subsubsection{Using SANA multiple network alignment on IID networks}
In the next experiment, SANA network aligner is used to produce real multiple network alignments \cite{sana}. SANA is an iterative aligner and we expected the quality of the alignments to increase with the number of iterations. IID mammalian networks were used for this experiment \cite{kotlyar2019iid}. We specifically used IID mammalian networks, which have a higher edge density compared to BioGRID networks, as the latter was found to have a lower number of edges than necessary based on information theory \cite{wang2022current}. To ensure meaningful results from iteration, we selected all the GO term annotated species within the IID mammalian networks, which included cow, dog, mouse, human, and rat. Within these networks, we created 30 real alignments for each combination of $k=3$, and $k=5$ species. The summary of the information of the utilized IID mammalian networks are shown in the table \ref{tab:network-summary}.

\begin{table}[htbp]
\centering
\caption{A summary of experimented IID mammalian networks}
\label{tab:network-summary}
\begin{tabular}{ccccc}
\hline
Nodes & Common Name & Official Name & Abbr & TaxID \\
\hline
18079 & Human & H. Sapiens & HS & 9606\\
17529 & Mouse & M. Musculus & MM & 10090\\
15740 & Rat & R. Norvegicus & RN & 10116\\
14512 & Dog & C. Familiaris & CF & 9615\\
14783 & Cow & B. Taurus & BT & 9913\\ 
\hline
\end{tabular}
\end{table}

The statistical significance of the exposed G measure was further validated by comparing it to the number of recovered orthologs within species. A higher number of recovered orthologs indicates a better alignment, providing additional objective evidence for the quality of the alignment. Therefore, the trend of the measures was compared to the trend of the recovered ortholog counts.

Generally, for the statistical exposed G, In addition to running it on all the go terms with $\lambda1$ below 100, we used sampling on 100, 1000, and 10000 go terms. Adding sampling options help to run the program in a very short amount of time. It is important to note that averaging was used to combine the $p$-values of different GO terms because here, we are using the statistical exposed G as a quality measure, and we didn’t want it to be affected by the number of GO terms in the networks. Whereas, if we wanted to assess the alignment only statistically, multiplying them would be close to the actual value since each GO term provides additional information on the statistical significance of the alignment. 

\section{Results}
\subsection{Exposed G and SGS score}
\subsubsection{Perfect self-alignment with controlled error rate}
\begin{figure}
\centering
\begin{minipage}{.45\textwidth}
  \centering
  \includegraphics[width=\linewidth]{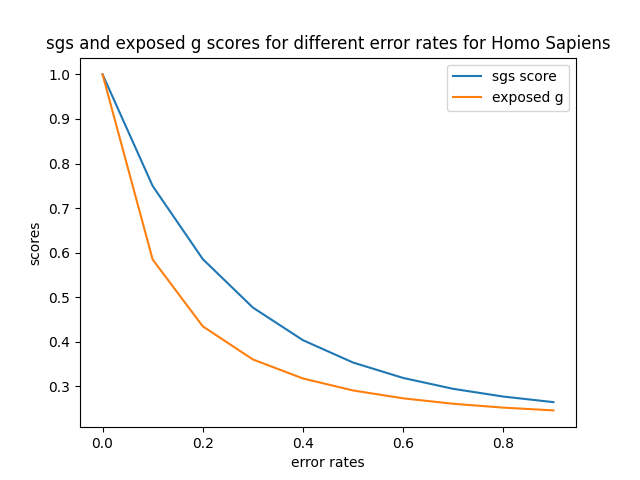}
  \includegraphics[width=\linewidth]{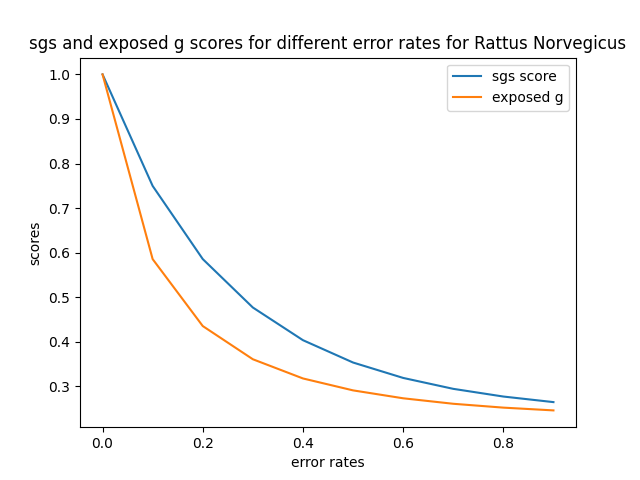}
\end{minipage}%
\hfill
\begin{minipage}{.45\textwidth}
  \centering
  \includegraphics[width=\linewidth]{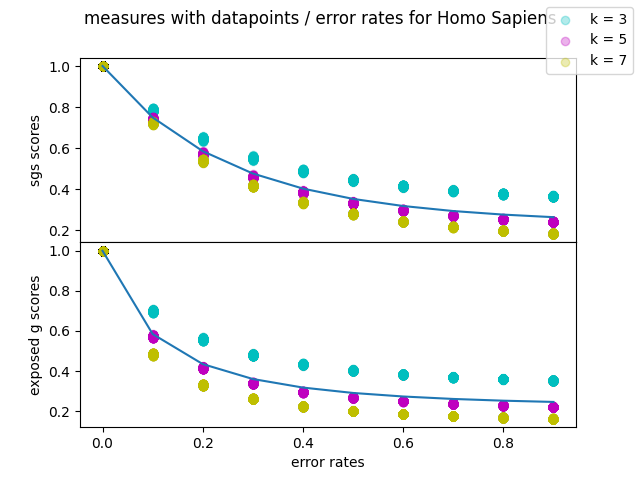}
  \includegraphics[width=\linewidth]{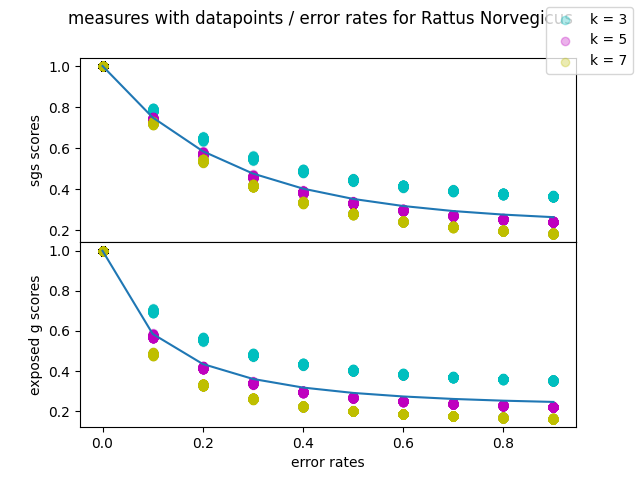}
\end{minipage}

\begin{minipage}{.45\textwidth}
  \centering
  \includegraphics[width=\linewidth]{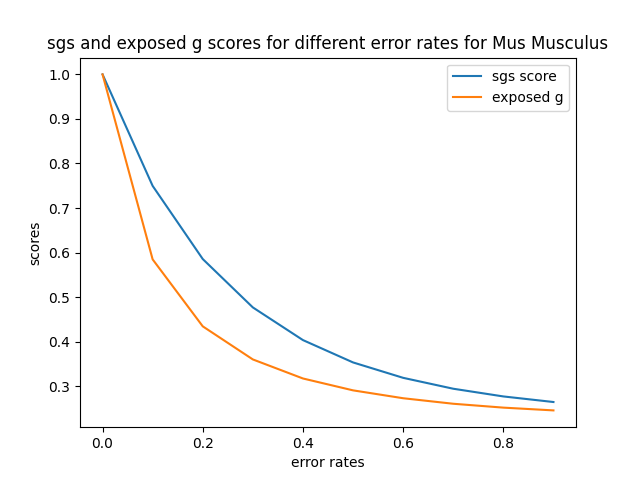}
  \includegraphics[width=\linewidth]{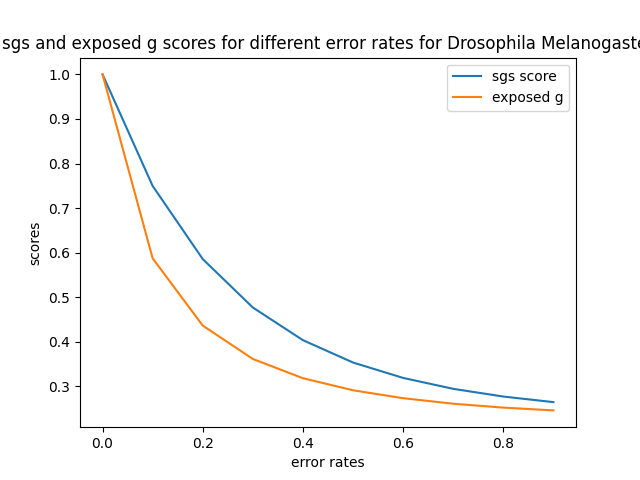}
\end{minipage}%
\hfill
\begin{minipage}{.45\textwidth}
  \centering
  \includegraphics[width=\linewidth]{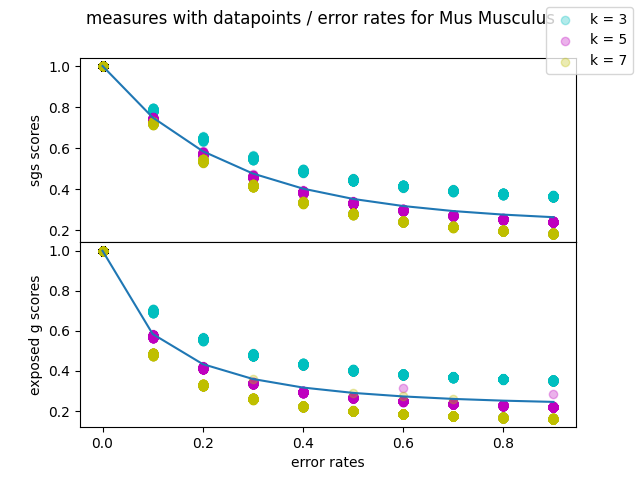}
  \includegraphics[width=\linewidth]{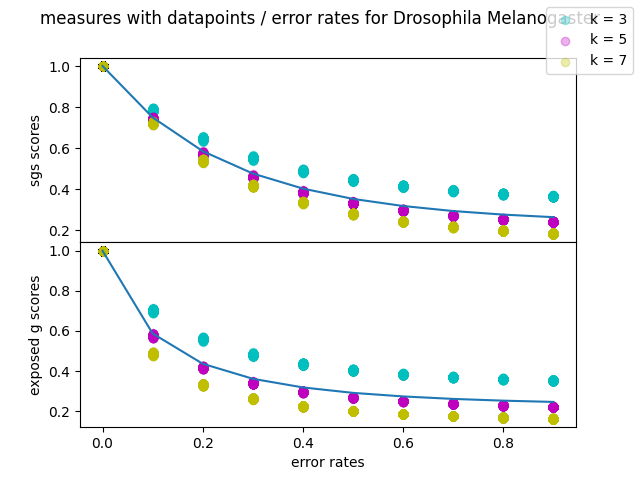}
\end{minipage}
\caption{The figures show the result of Self-alignment experiment for BioGrid networks. Each row corresponds to one species; from above: human, rat, mouse, and fruit fly. The right column show every single datapoint on the plot separated for exposed G and SGS, while the left column compares the trend of SGS and exposed G.}
\label{fig:self-alignmentresult}
\end{figure}

Figure \ref{fig:self-alignmentresult} display the results of the perfect self-alignment with a controlled error rate for both the exposed G and SGS score. Each circle in the right figures represents a single data point. As depicted in the figures, the data points are concentrated on a single point for every $k$ and error rate, indicating that the measures are reliable in distinguishing the quality of alignments. 

Additionally, both SGS and exposed g measures are monotonically decreasing with error rate. The reason that higher values of $k$ cause the score of the measures to decrease is that introducing the same error rate for a larger number of networks causes a higher percentage of perfect towers to be damaged.

\subsubsection{Using SANA multiple network alignment on IID networks}
\begin{figure}[htbp]
  \centering
  \begin{subfigure}
    \centering
    \includegraphics[width=\linewidth]{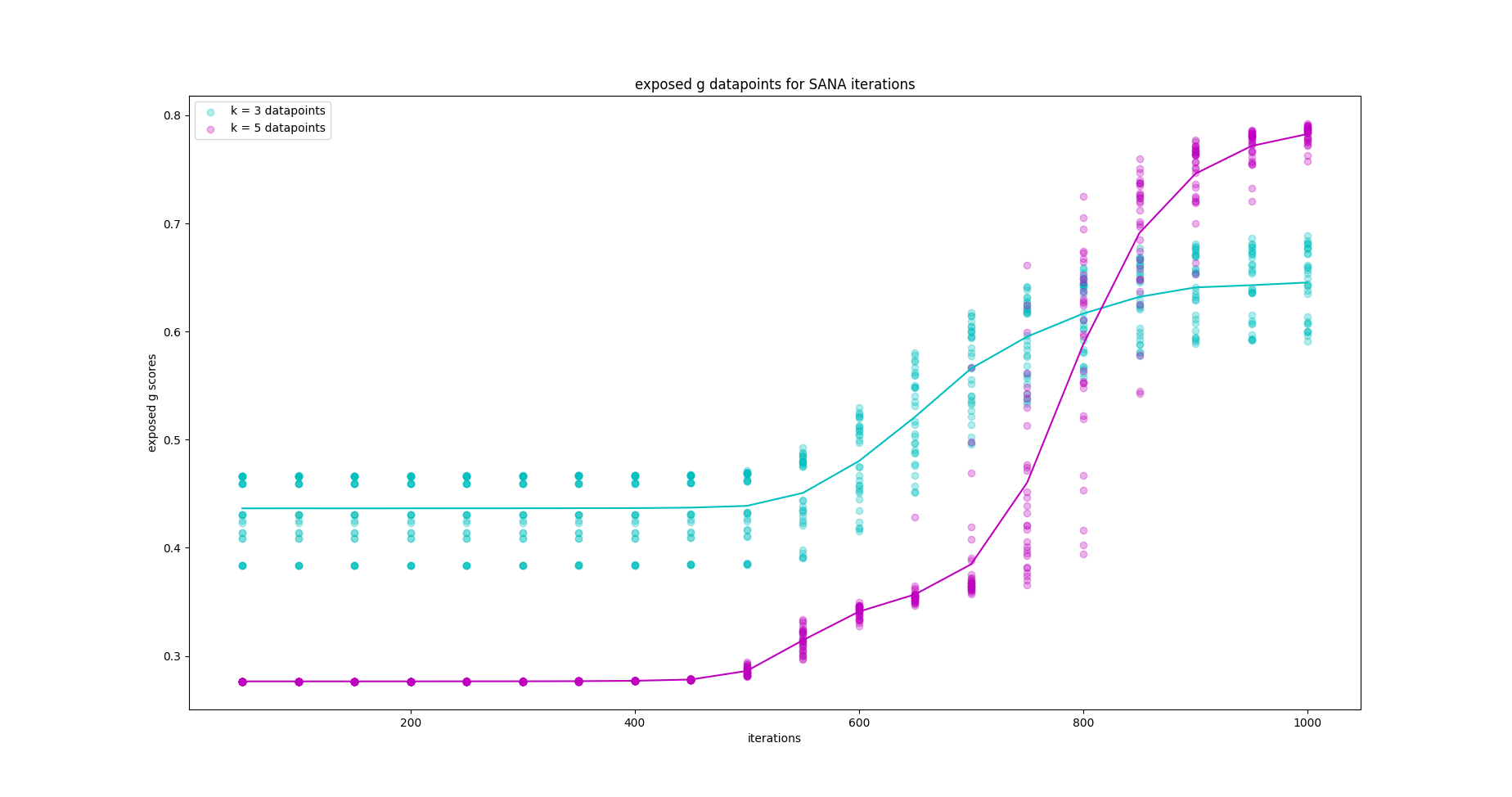}
  \end{subfigure}
  
  \begin{subfigure}
    \centering
    \includegraphics[width=\linewidth]{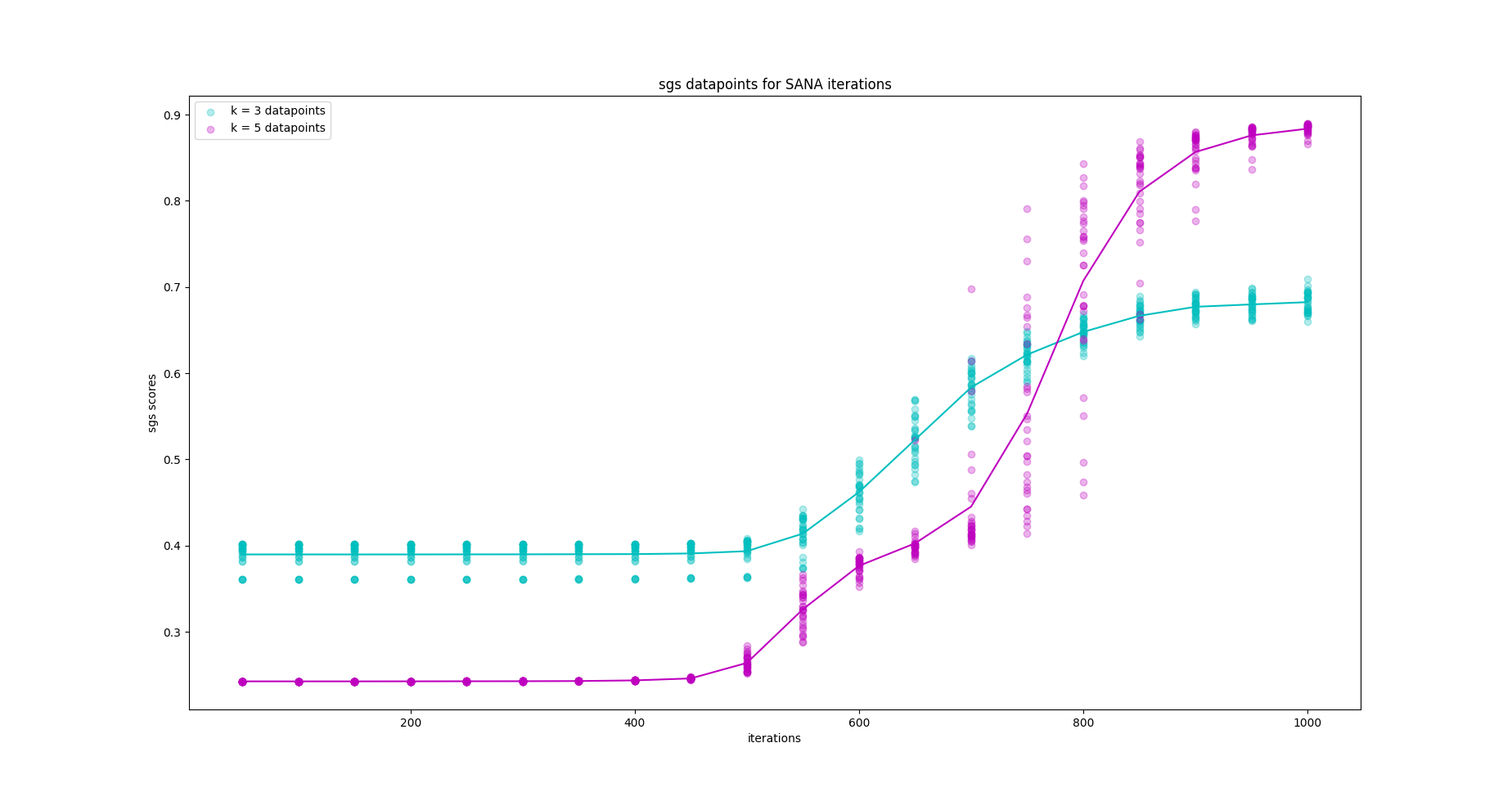}
  \end{subfigure}
  \caption{The plots show the result of exposed G and SGS, after running SANA for 1000 iteration on IID mammalian networks. The pink curves are related to 5 networks and the blue lines to 3 networks. The first figure is the result of exposed G and the second one is for SGS.}
  \label{fig:realrealexpsgs}
\end{figure}

Figure \ref{fig:realrealexpsgs} show the result of SGS and exposed G for SANA multiple network aligner on IID networks. The curves for both SGS score and exposed G score show a monotonically increasing pattern with iteration, as illustrated in the figures. The curves for SGS score and exposed G score produce very close absolute values, as shown in the first figure. Exposed G score is capable of distinguishing alignments with a different set of species for different alignments for $k=3$, while for $k=5$, all alignments for both measures get close final results. To combine the $p$-values of the GO terms for a holistic $p$-value, we used averaging.

\subsection{Statistical Exposed G}
\subsubsection{Perfect self-alignment with controlled error rate}
\begin{figure}[htbp]
  \centering
  \begin{subfigure}
    \centering
    \includegraphics[width=0.7\linewidth]{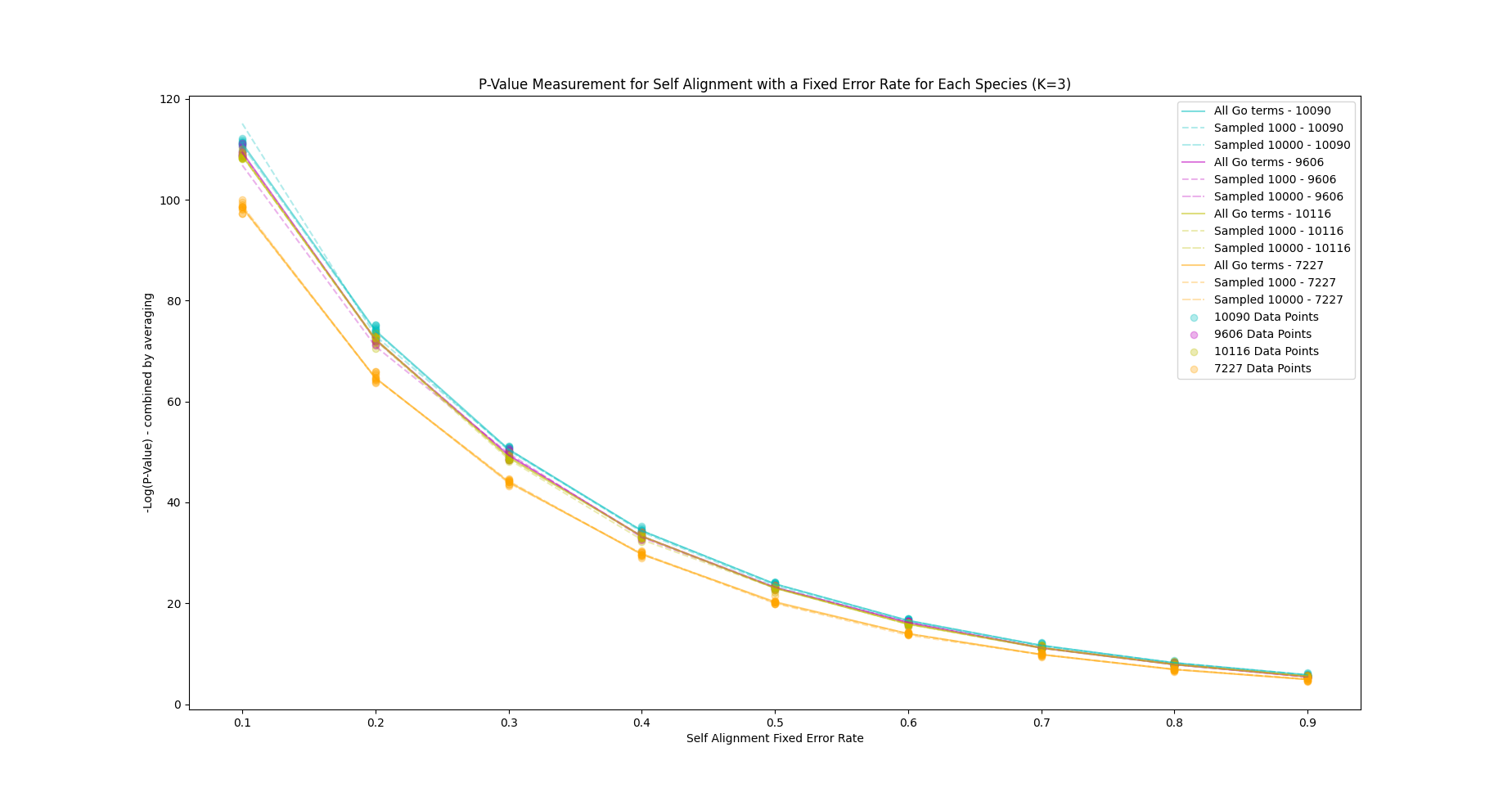}
  \end{subfigure}
  
  \begin{subfigure}
    \centering
    \includegraphics[width=0.7\linewidth]{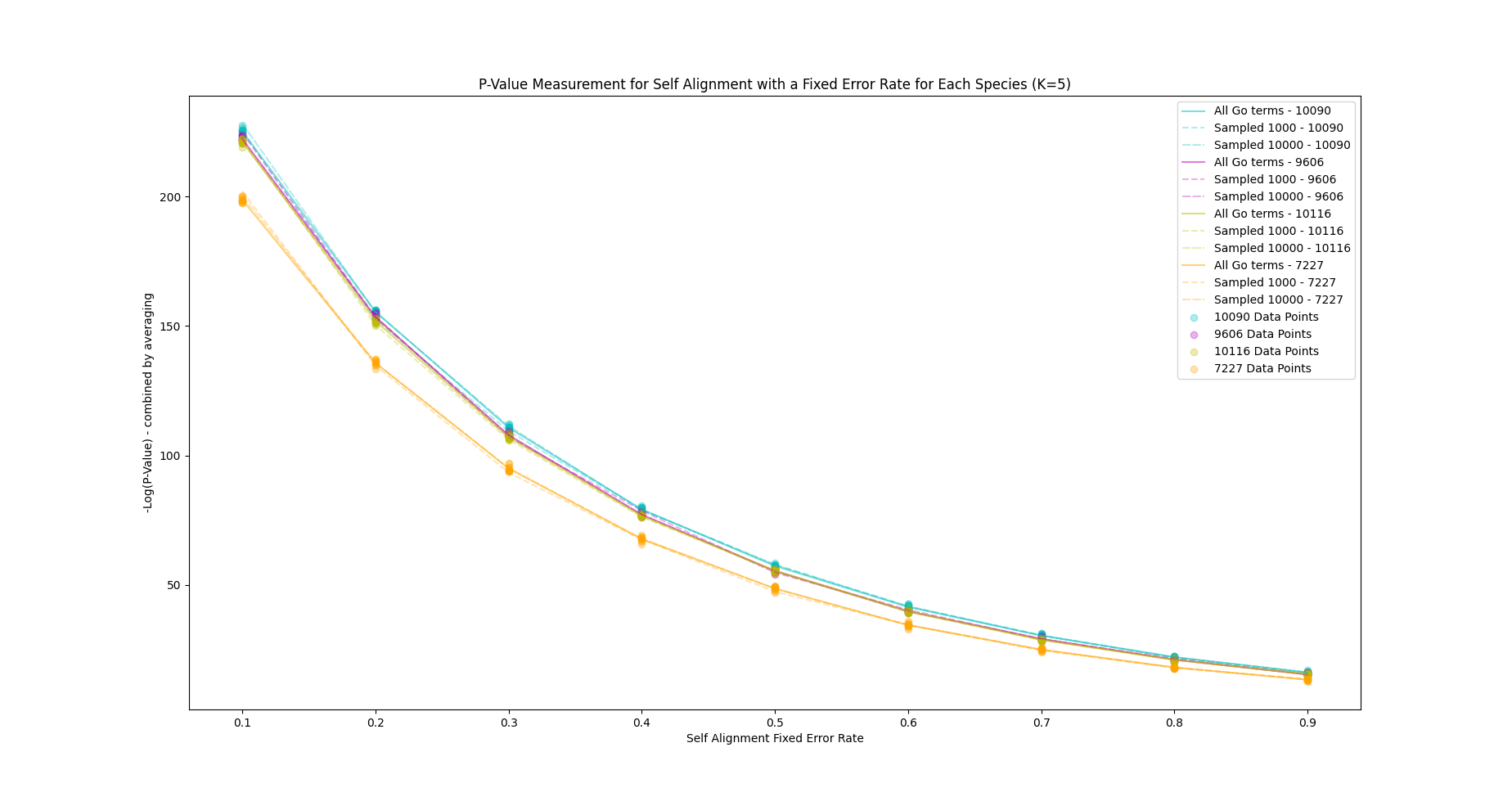}
  \end{subfigure}

  \begin{subfigure}
    \centering
    \includegraphics[width=0.7\linewidth]{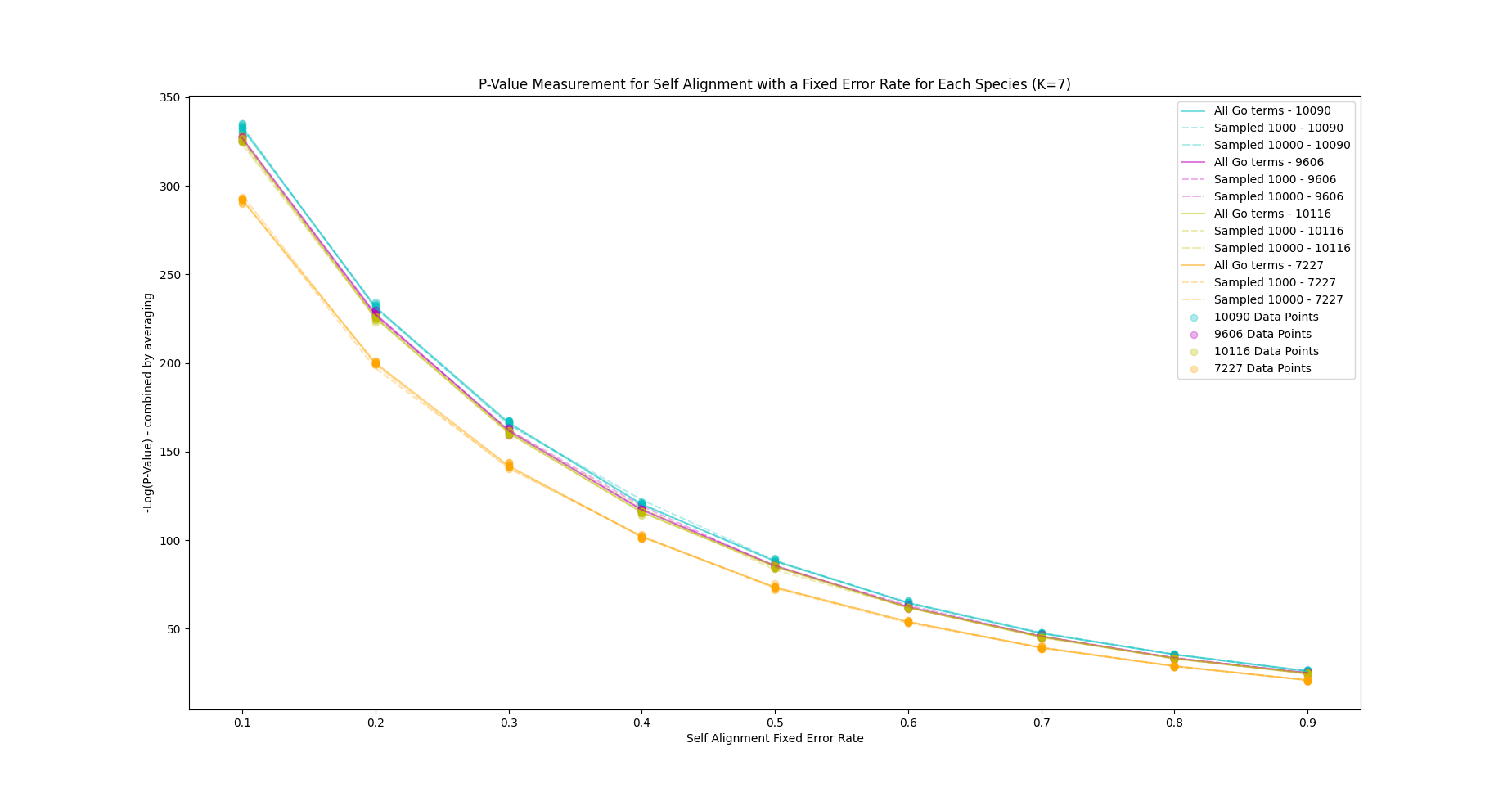}
  \end{subfigure}
  \caption{The plots show the result of statistical exposed G for self-alignments between from above: 3, 5, and 7 IID networks. The $p$-values start lower by increasing the number of networks, as it's more significant to generate high-quality alignments for a larger number of networks. The scale uses natural logarithms.}
  \label{fig:selfp-values}
\end{figure}

Figure \ref{fig:selfp-values} show the result of statistical exposed G on self-alignments. We tested the statistical $p$-value measure on self alignments with a fixed error rate. Fruit Fly ($TaxID=7227$) network is smaller than the other species and therefore, even when it reaches zero error rate, the $p$-value is less significant than other species. The other measures (Exposed G and SGS) were unable to detect these differences, but statistical exposed g offers a tool that can be used for more global comparisons. 
The second plot shows the result of the $p$-value alignment for self-alignments with a fixed error rate between 5 networks. This plot shows that reaching an error rate with $K=5$ is more significant than reaching the same error rate for $K=3$. The last plot shows the results for 7 networks. The same trend is visible here, and again, the $p$-values are lower. 

\subsubsection{Using SANA multiple network alignment on IID networks}
\begin{figure}[htbp]
  \centering
  \includegraphics[width=0.9\textwidth]{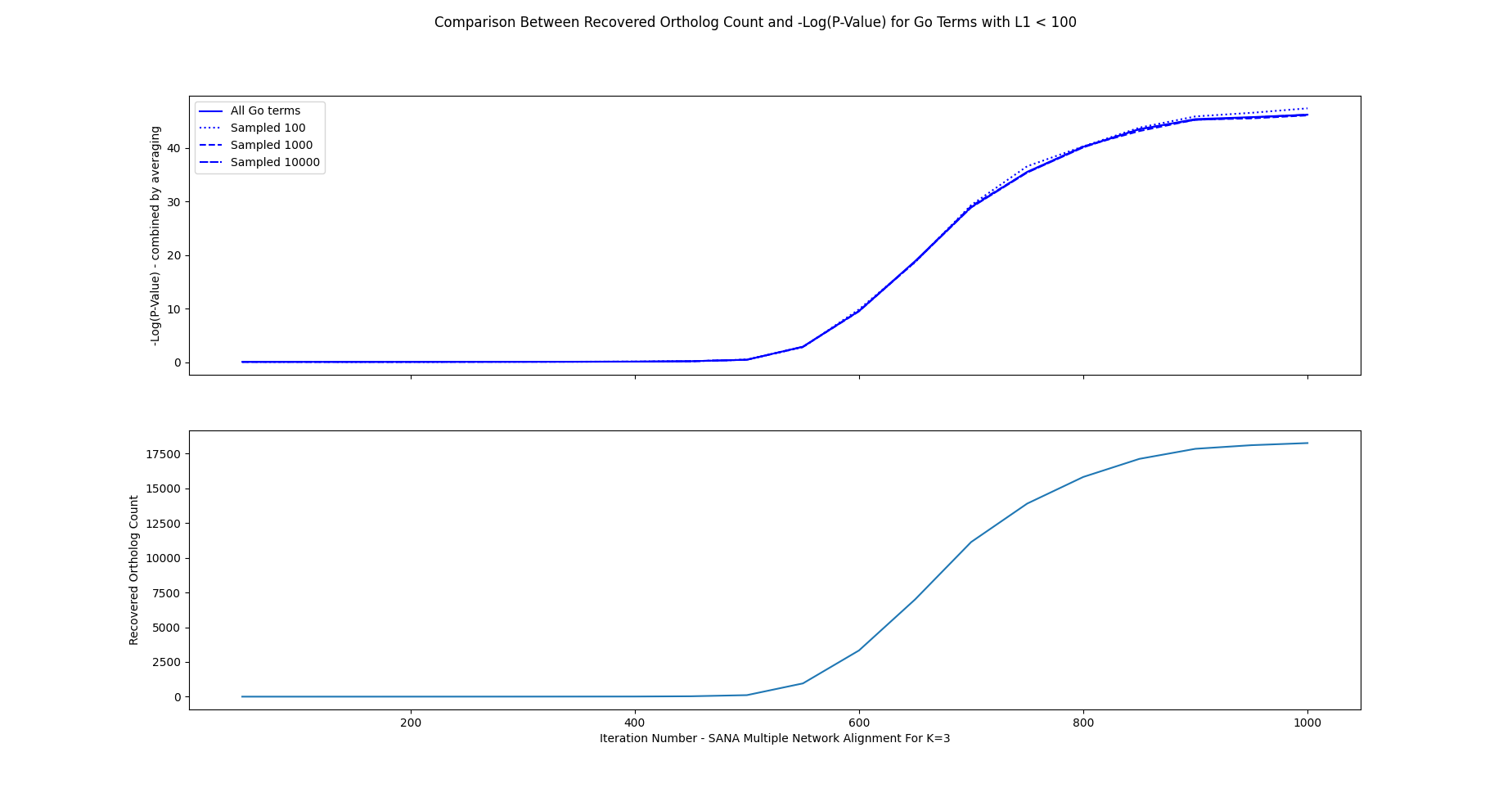}
  \caption{This figure shows the result of statistical exposed G for different iterations of SANA multiple network aligner for 3 networks. The second plot, shows the number of recovered Ortholog count for the same alignments. Both curves match which shows the statistical exposed G is a promising indicator of quality.}
  \label{fig:pvalreal-3}
\end{figure}

\begin{figure}[htbp]
  \centering
  \includegraphics[width=0.9\textwidth]{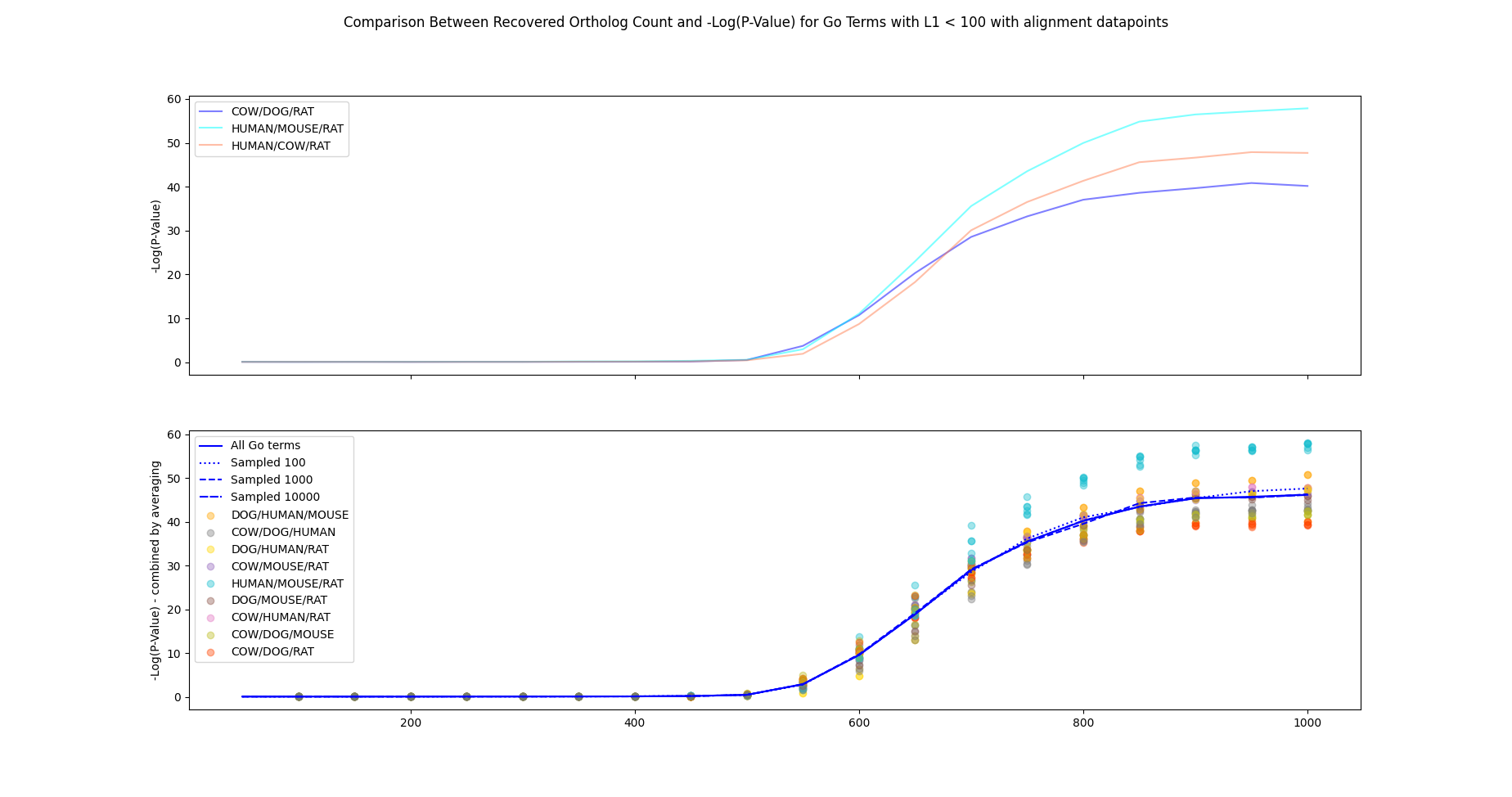}
  \caption{This figure shows the differences of species and the datapoints within the result of statistical exposed G for different iterations of SANA multiple network aligner for 3 networks. More complicated sets of species land on lower $p$-values.}
  \label{fig:pvalrealdiffs}
\end{figure}

For these experiments, we observed that the trend in the figure \ref{fig:pvalreal-3}, which shows the results for 3 networks, matches exactly with the recovered ortholog count, which provides further evidence for the legitimacy of the measures. Additionally, in figure \ref{fig:pvalrealdiffs} we observed that in species like mouse and human, where the PPI network is more complete, the final score is more statistically significant. The more difficult alignments with lower $p$-values improved more slowly around the iteration 500. 

\begin{figure}[htbp]
  \centering
  \includegraphics[width=0.9\textwidth]{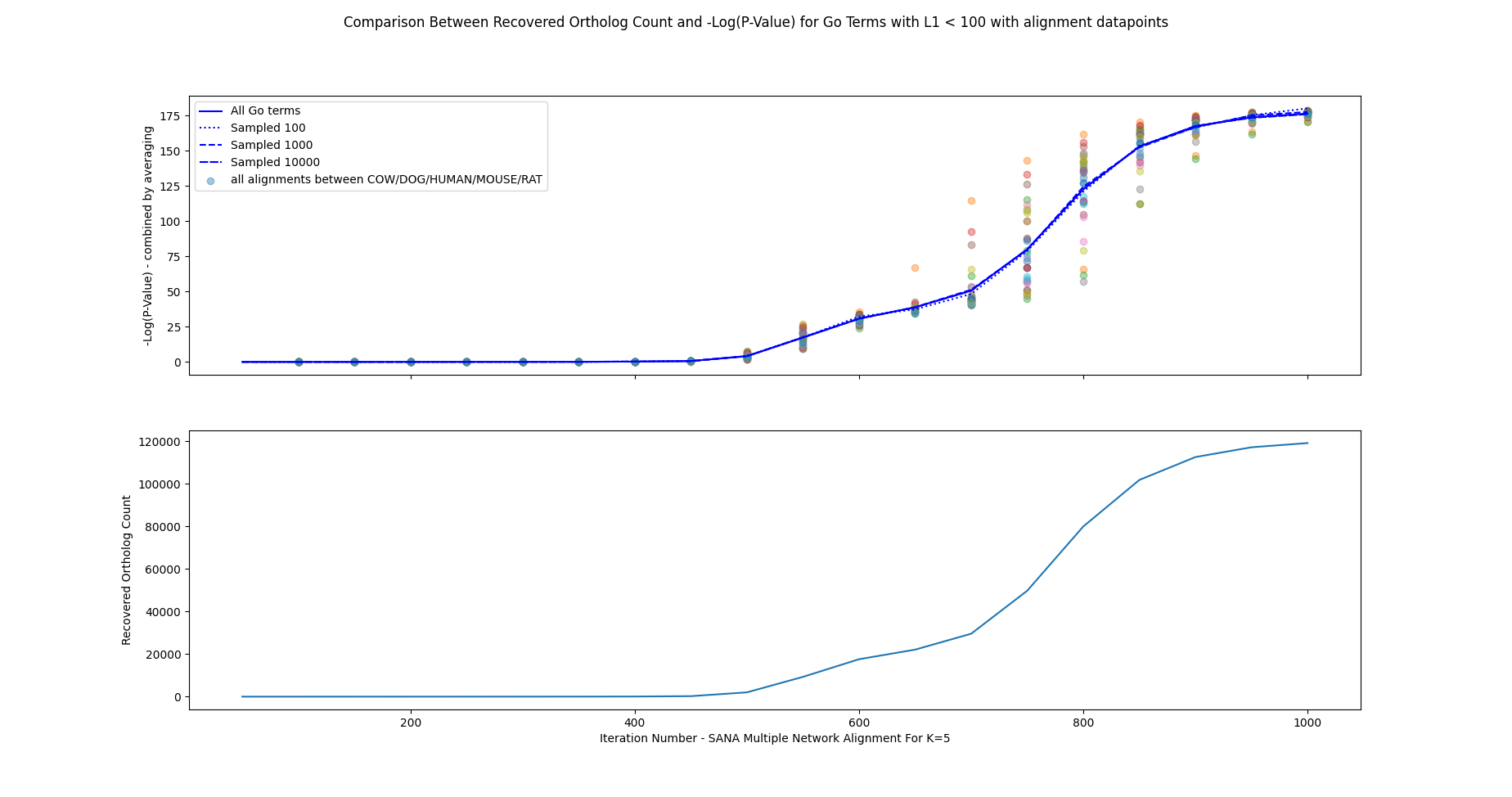}
  \caption{This figure shows the result of statistical exposed G for different iterations of SANA multiple network aligner for 5 networks. The second plot, shows the number of recovered Ortholog count for the same alignments. Both curves match which shows the statistical exposed G is a promising indicator of quality.}
  \label{fig:pvalreal-5}
\end{figure}

In figure \ref{fig:pvalreal-5}, we observed that for $k=5$, the ortholog recovered count exactly matches with the log $p$-value curve. Additionally, reaching good alignments is more statistically significant than in $k=3$. Another noteworthy difference is that the data points become very close as the iterations go by. This shows that for similar combinations of species, almost the same $p$-value is reached, although the process of the alignment is different for different alignments.

\section{Discussion}
Our evaluation of SGS, exposed G yielded promising results. Firstly, we observed a perfect match between the expected trend and our measures when we used a perfect self-alignment with a controlled error rate. Secondly, we demonstrated that our measures had a perfect match with the recovered ortholog count and with each other when we used the SANA multiple network aligner and allowed it to run through 1000 iterations on IID mammalian networks. 

Furthermore, our $p$-value measure was fully supported by our proposed measures and the recovered Ortholog counts. The $p$-value measure was validated by testing it against empirically generating random alignments and matching the empirical $p$-value against our calculated $p$-value. This measure was found to have a perfect curve very close to the empirical $p$-value. We also validated statistical exposed G by observing that it sums to one when combined for all the possibilities. 

\subsection{Limitations}
Despite these promising results, there are several limitations to our study that must be acknowledged. Firstly, we combined scores for different GO terms using averaging, which may have overlooked their inter-dependencies. Secondly, to resolve the adverse effect of the large number of log-summations required on the precision, we restricted our analysis to GO terms with a $\lambda1$ value of 100 or less. These GO terms accounted for the majority of available GO terms in the network. Additionally, we set the size of the shadow network to be equal to the largest network and did not allow for outer matches, as setting a maximum number of extra shadow nodes would have had a significant impact on the $p$-value, which should reflect the quality of the alignment as indicated by the Exposed G measure.

\subsection{Future Work}
In the future, we plan to further enhance our approach by computing the p-values for Squared GO Score (SGS) and a full combinatorial p-value. We also plan to use Empirical Brown's Method to account for the inter-dependencies between the GO terms. Additionally, we want to further test our measures on BioGrid, to learn how they perform on unbalanced data. 

\subsection{Conclusion}
In conclusion, our proposed measures have shown promise in assessing the quality of multiple network alignment in PPI networks. The ability to evaluate the statistical significance of a multiple network alignment is particularly important for the analysis of large-scale biological networks. Our work has potential applications in predicting gene function and understanding the functional relationships between proteins. 

\section{Appendix}
\subsection{Optimization of statistical Exposed G}

The numerator of the statistical measure is computationally intensive, with a computational complexity of $O((n_0-n_1)^2 * (exposed G - \lambda1)^2 * k)$, where $n_0-n_1$ represents the number of extra shadow nodes used. To address this, we utilized a hashmap memory for log factorials and different functions of the numerator. We also observed that different parts of the summations converged before the completion of the summation. However, computation for one $E(x)$ took approximately one second, which is time-consuming as $E(x)$ needs to be calculated for all values below exposed G. Furthermore, as this calculation needs to be performed for all GO terms, which number around 20,000, this process is computationally expensive but easily paralleled.

\begin{figure}[htbp]
  \centering
  \includegraphics[width=0.9\textwidth]{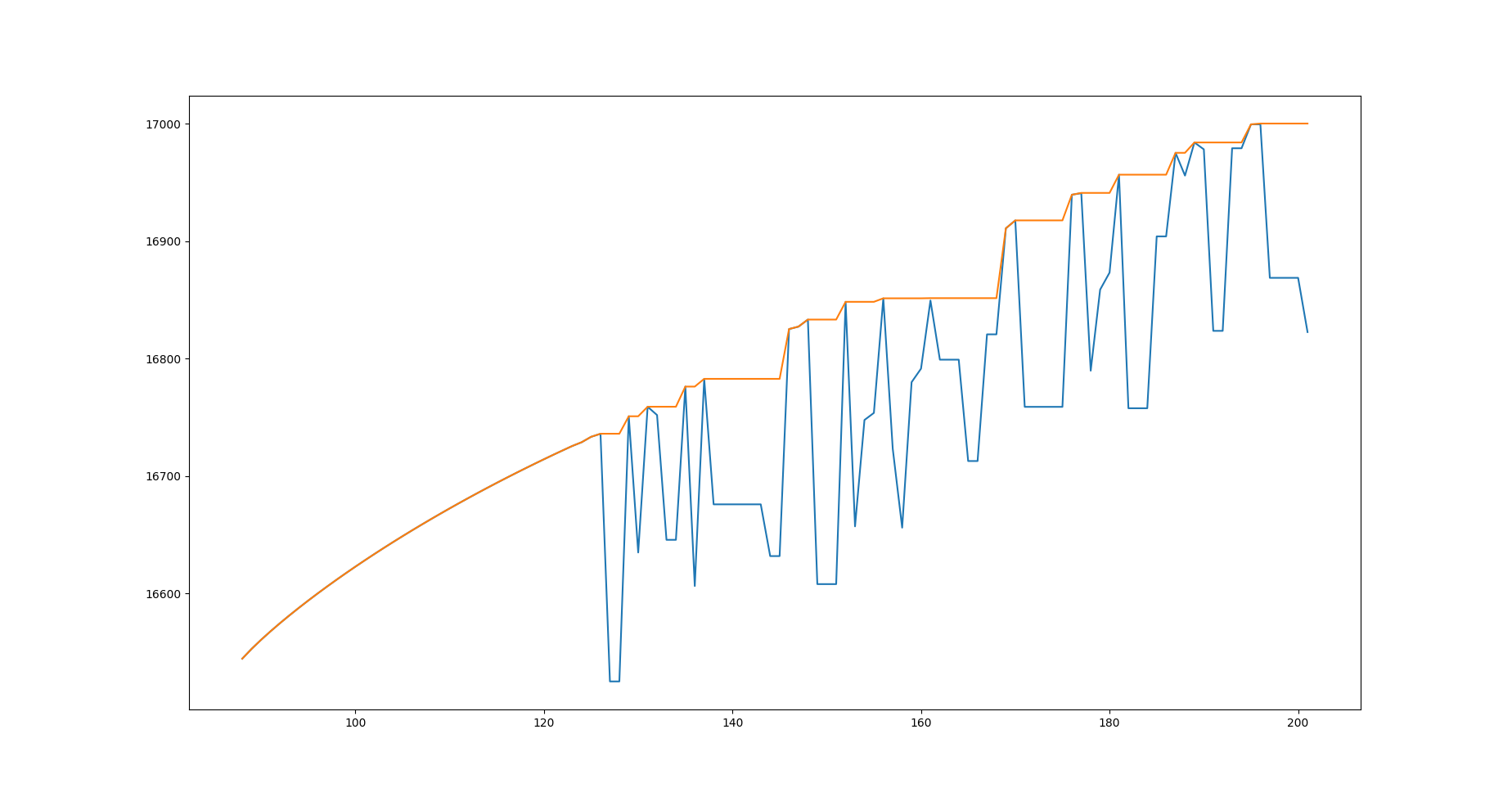}
  \caption{This figure shows that the general trend of E(x) for a point estimate of an exposed G is generally increasing and is very close to the cumulative calculation if we use the peaks.}
  \label{fig:cumulative}
\end{figure}

The optimization is optional for the cases where parallelism isn't used and the $p$-value is needed to be produced on all of the GO terms. To reduce the need for calculating the measure for all $x$ values between the $\lambda_1$ and the exposed G, we plotted $E(x)$ and cumulative $E(x)$ as viewed in figure \ref{fig:cumulative}. Our analysis showed that the $E(x)$ curve increases generally, with the maximum points and cumulative $E(x)$ points being very close as shown in the figure below. To accurately reflect the quality of statistical exposed G, we only required a correctly calculated increasing curve by exposed G. Therefore, instead of calculating everything, we started with the exposed G and worked backward until we reached the previous peak. We then returned the current value. This returned the result within a few runs of $E(x)$. We used the derivative of the function as an approximation of the error, and found the error to be negligible. We tested this approach on both random synthesized network states and real alignments, and found it reduced the time required from 5 minutes and 41 seconds to 18 seconds for real mammalian networks with exposed G in the middle of the allowed range for a single GO term.

We experimented with different values of shadow holes and observed that using shadow nodes reduced the $p$-value by a significant factor, but did not indicate that we have a better alignment. This means that if two people used different numbers of maximum extra shadow nodes for the same multiple network alignment, they would receive different results. Consequently, we used the biggest network as the alignment base to ensure that the $p$-value was only affected by the exposed G as the quality indicator. The use of the zero extra shadow holes counterpart for the alignments could only be problematic for GO terms that are poorly aligned enough to exceed the number of nodes in the biggest network. However, this was only observed in a small number of GO terms with an average above $N/k$ instances per network, which were common and uninformative GO terms. For these GO terms, we safely assumed the value of 1 for the $p$-value, as the holistic $p$-value is a combination of a large number of GO term $p$-values. Furthermore, using logarithmic summation and subtraction can significantly affect the precision. By removing the shadow holes, the time complexity dropped to $O((exposed G-\lambda1)^2)$, and only two series of summation were required. Nonetheless, we used fixed-precision libraries with at least 60 decimal point precision to obtain valid results. It is important to note that setting extra shadow nodes is accounted in the equations and the codes and can be used but it's not recommended. 

\begin{figure}[htbp]
  \centering
  \includegraphics[width=0.9\textwidth]{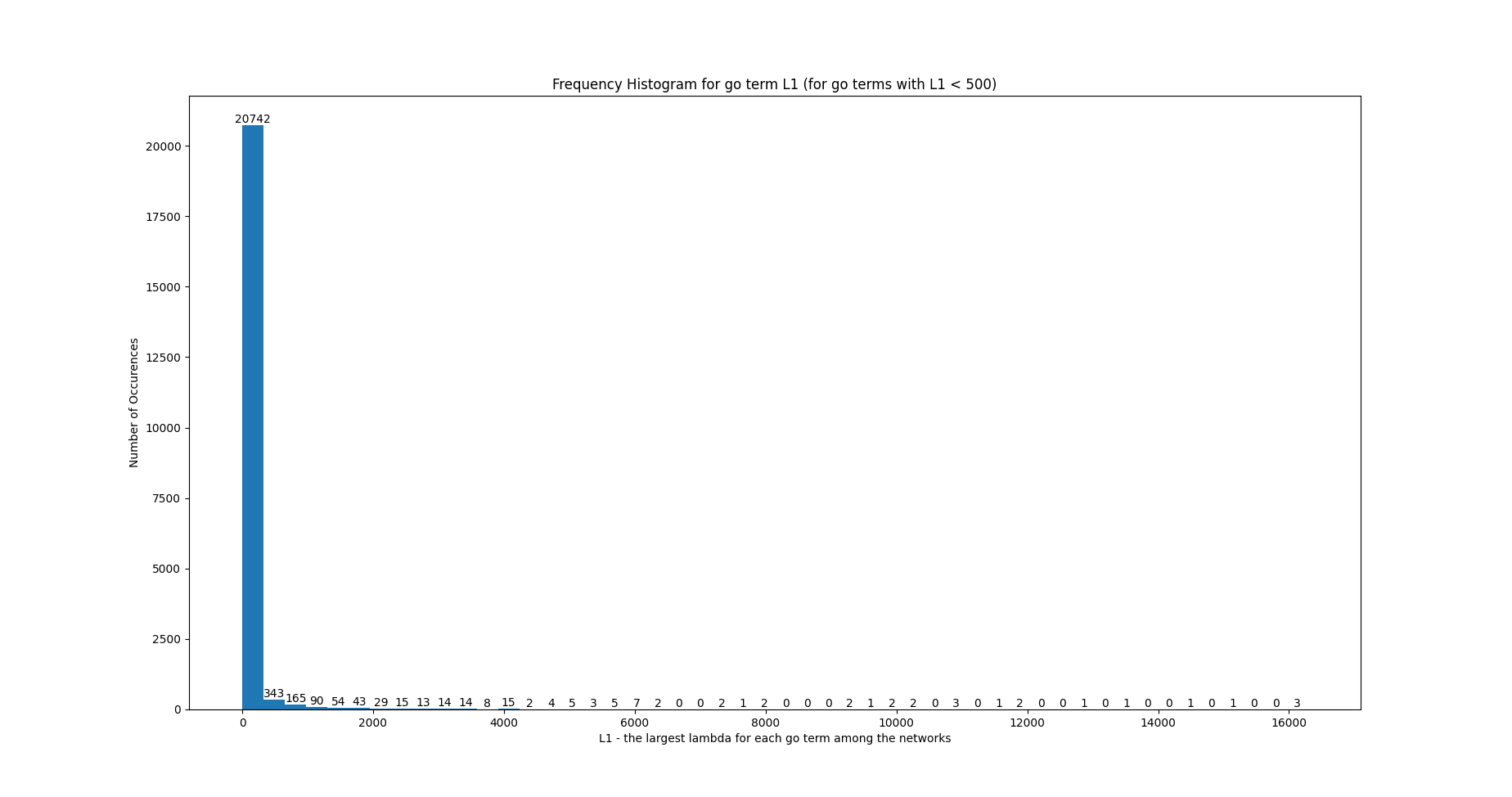}
  \caption{This figure plots the frequency histogram of $\lambda1$s for all GO terms. It shows that the majority of the GO terms have $\lambda1$s below 100.}
  \label{fig:freqs}
\end{figure}

For smaller exposed g values a shorter summation and subsequently a smaller precision was required. For bigger exposed g which was the case for the networks with big lambdas higher precision was needed. Using higher precision than 300 decimal points slowed the process too much. Fortunately, almost all of the go terms had lambdas smaller than 100 as demonstrated in the figure \ref{fig:freqs}. It is important to note that the smaller the $\lambda1$ of a go term is, it is more likely that the go term is more informative and more specified for a certain function. Which means go terms with smaller lambdas are representative and should contribute to the final result. Using go terms with higher than a hundred lambdas, in most cases, had big errors due to the summations. As a result, we use go terms with lower than a hundred $\lambda1$ for obtaining the final result. It is also possible to use sampling options and only look at a specified number of go terms, but the result for all the go terms with $\lambda1$ below 100 is usually calculated within an hour. 

\begin{figure}[htbp]
  \centering
  \includegraphics[width=0.9\textwidth]{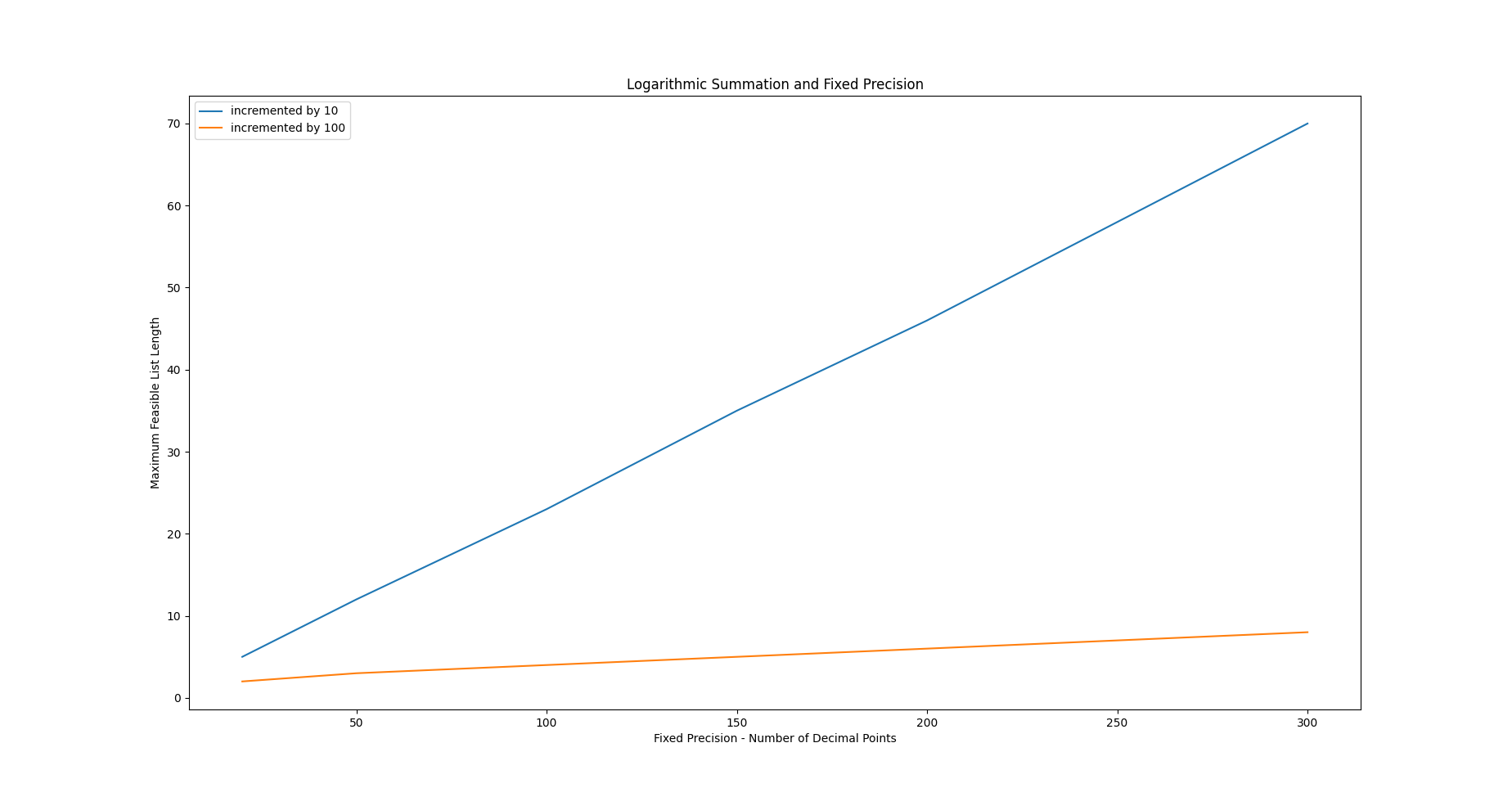}
  \caption{This figure plots the number of log-summations done before the precision falls. The blue line shows the result when the numbers are distanced by 10, and the orange one shows it for 100.}
  \label{fig:precisions}
\end{figure}

In order to see how logarithmic summation and subtraction is affected by a fixed precision we designed an experiment. We created a list by starting from a value close to our denominator and incrementing the value by 10 or 100 to get the next element. We used logarithmic summation on this list and then logarithmic subtraction to remove all the elements except for the first one. By increasing the length of this list, more precision was required to get results. We noticed that the precision is either close to perfect or completely out, which was close to what we were observing in the actual runs. Therefore, we wanted to see how many elements could be handled before we reach the precision drop for different fixed precisions. The precision wasn’t affected by the first value that we chose, it was only affected by the fixed precision and the value by which we incremented. As shown in figure \ref{fig:precisions}, the closer the values, more elements could be handled with high precision. However, still even with 300 decimal points below 70 summations could be done for a difference of 10.

As the plot was linear we were almost able to derived that $2/incrementValue$ was a good approximation of the slope. We validated this by checking it with an increment value of 1 and shuffling the lists. We calculated the mean of the differences that we encountered in the real alignments which had the increment value of around 3, so we used 0.66 as the slope to calculate the fixed precision that we need. Using this, the precision is allocated dynamically for a minimum value of 60 and a maximum value of 300, and using these equations to obtain the number. The calculations are even more precise and faster for better alignments which ensures that good alignments are easily distinguished from each other even in very low $p$-values. The figure also shows why we can’t reach the required precision for GO terms with a big $\lambda1$. The number of logarithmic summations needed is $exposed g - 1 + 1$ for each GO Term which is in most cases too big for these go terms. 

Overall, the optimizations allowed us to calculate the $p$-values for the majority of the GO terms within an alignment with a high precision and a reasonable time.

\bibliographystyle{elsarticle-num}

\section*{References}

\bibliography{paper}


\end{document}